\documentstyle[prl,aps]{revtex}
\begin{document}
\draft
\title{Measurements of Intensity Distributions 
in the Approach to Localization}

\author{M. Stoytchev and A. Z. Genack}
\address{Dept. of Physics, Queens College of CUNY, 
65-30 Kissena Blvd., Flushing, NY 11367}

\maketitle

\begin{center}
(April 10, 1998)
\end{center}

\begin{abstract}
We report measurements of intensity distributions of transmitted 
microwave radiation in quasi-1D samples with lengths $L$ as large 
as the localization length $\xi$. In contrast to negative exponential 
statistics found in the diffusive limit, the distribution falls as a 
stretched exponential of power $1/2$ and becomes nearly log-normal as 
$L$ approaches $\xi$. We confirm the relation between the moments and 
full distributions of the intensity and total transmission obtained by 
Kogan and Kaveh using random matrix theory. Good agreement is found 
when this relationship is used to compare measurements of intensity 
distribution in strongly absorbing samples with lengths as large as 
$\xi$ with calculations by Brouwer of the transmission distribution. 
The variances of the measured distributions increase superlinearly near 
the localization threshold confirming the coexistence of localization 
and absorption and the suitability of these variances as measures of 
the closeness to the localization threshold.
\end{abstract}
	As the localization threshold is approached, fluctuations in 
key transmittance quantities become as large as the ensemble average 
values of these quantities. Thus a comprehensive description of 
mesoscopic transport should provide the full distribution of 
transmittance quantities and the relationships between them. \cite{Shap86} 
In order of increasing spatial averaging, key transmittance quantities 
are the intensity, corresponding to the transmission coefficient 
$T_{ab}$ for incident mode $a$ and outgoing mode $b$, the total 
transmission, $T_a = \sum_b T_{ab}$, and the transmittance, 
$T = \sum_{ab} T_{ab}$. In the absence of absorption, the transmittance 
equals the dimensionless conductance $g = N\ell/L$, where $N$ is the 
number of transverse modes and $\ell$ is the transport mean free path. 
\cite{Land70}  Localization in quasi-1D samples is achieved when $g = 1$ 
at $L = \xi = N\ell$. In this Letter, we focus on the intensity 
distribution, which is the key distribution in statistical optics. 
\cite{Goodman} We demonstrate its relationship to the distribution of 
total transmission, find the scaling of the variance of the intensity 
and total transmission up to $L = \xi$, and determine the extent to 
which absorption influences localization.

In the diffusive limit, the degree of long-range intensity correlation is 
small and the intensity distribution is well approximated by the Rayleigh 
distribution. \cite{Goodman,Gar89,Dainty} For polarized detection, this 
corresponds to negative exponential statistics, $P(s_{ab}) = exp(-s_{ab})$, 
where $s_{ab} = T_{ab}/<T_{ab}>$ is the intensity normalized to its 
ensemble average value. In previous work, deviations from negative 
exponential behavior have been observed and ascribed to long-range 
intensity correlation \cite{Gar89,Gen93,Shnerb}. In these studies, 
fluctuations as large as $s_{ab} \sim 10 \sim g$ were observed. In the 
present work, fluctuations as large as fifty times $g$ are observed in 
samples with lengths $L \sim \xi$. 

The intensity distribution is studied in a quasi-1D geometry, which is 
equivalent to the electronic case of a thin wire. Thouless argued that, 
the level width $\delta \nu$ in a wire at $T = 0$ should become smaller 
than the level spacing $\Delta \nu$ since $\delta \nu$ is proportional to 
the inverse of the travel time and so falls as $1/L^2$ in the diffusive 
limit, whereas $\Delta \nu$ is the inverse of the density of states and 
so falls as $1/L$.  As a result, the modes in adjacent sections of the 
wire should not overlap, and electrons should become localized. 
\cite{Thls77} 

The question arises as to whether radiation can be localized in the 
presence of absorption. In this case, the level width falls as $1/L$, 
just as the level spacing does. In previous measurements in absorbing 
samples, the variance of the normalized total transmission, 
$<s_{ab}> = T_a/<T_a>$, $var(s_a)$, was found to scale sublinearly 
with $L$. \cite{ms97} If the attenuation length due to absorption, 
$L_a$, serves as a cutoff length for localization \cite{John84,Gen91} 
then $var(s_a)$ would approach an asymptotic limit as $L$ increases. 
On the other hand, if localization can be achieved in absorbing samples, 
then $var(s_a)$, which is essentially the degree of correlation in the 
intensity of different outgoing modes, should increase superlinearly as 
$L$ approaches $\xi$. This might occur, since the wave remains 
temporally coherent in the presence of absorption. \cite{Weav93,Yos94} 
Weaver has shown in a 2-D simulation that the introduction of absorption 
does not disrupt the spatial localization of acoustic waves in closed 
systems, though the overall energy decreases exponentially with time. 
\cite{Weav93} In recent calculations, Brouwer found that for diffusive 
waves the prefactor multiplying $L/\xi$ in the expression for 
$var(s_a)$ drops from $\large \frac{2}{3}\normalsize$ to 
$\large \frac{1}{2}\normalsize$ as the ratio $L/L_a$ increases. 
\cite{Brw97} The behavior of this quantity, however, was only considered 
for lengths considerably less than the localization length. 

Here we report measurements of intensity transmitted through random 
waveguides with $L \leq \xi$, but $ >L_a$. We expect that modes in 
this sample are completely mixed and the degree of intensity correlation 
between different modes is constant. Wave propagation in this sample 
should therefore be described by random matrix theory (RMT). 
\cite{Stone91} Recently, Kogan and Kaveh used RMT to obtain a relationship 
between the moments of intensity and total transmission in nonabsorbing 
quasi-1D samples. \cite{Kog95} They find, 
\begin{equation}
\label{moments}
<\!\!s_{ab}^n\!\!> = n!<\!\!s_a^n\!\!>,
\end{equation}
\noindent
This leads to a relationship between the distributions of intensity and 
total transmission, \cite{Kog95,Kog93} 
\begin{equation}
\label{psab}
P(s_{ab}) = \int_{0}^{\infty}\frac{ds_a}{s_a}P(s_a)exp(-\frac{s_{ab}}{s_a}).
\end{equation}
\noindent
Since the distribution of total transmission can be calculated from the 
distribution of the eigenvalues of the transmission matrix, 
\cite{Kog95,vR95,vLan96} these relations provide a basis for calculating 
the intensity moments and distributions from RMT. 

The intensity distribution is obtained from measurements of the field 
transmitted through an ensemble of random configurations of 1.27 cm-diam. 
polystyrene spheres inside a copper tube. Tubes with diameters $d = 5.0$ 
and 7.5 cm and lengths up to 520 cm are used. The samples have filling 
fractions of 0.52 and 0.55 for $d = 5.0$ and 7.5 cm, respectively. The 
sample tube is rotated between successive measurements to produce new 
configurations of scatterers. At least two thousand sample configurations 
were used for each distribution. Field spectra were taken from 16.8 to 
17.8 GHz in steps of 1 MHz using a Hewlett-Packard 8722C network analyzer. 
The radiation is coupled into and out of the sample by 0.4 cm wire antennas 
placed 0.5 cm from the ends of the sample. In order to ensure that the 
distributions were not distorted by noise, it was necessary to use an 
amplifier with an output power of $40$ W for samples with lengths greater 
than 200 cm so that the average intensity was at least three hundred times 
the noise.  In the frequency range of the measurements, $\ell \sim 5$ cm 
\cite{Gen93a}. A fit of measurements of the field autocorrelation function 
with frequency shift to theory \cite{Shap-pc} gives $L_a = 34 \pm 2$ cm 
and $D = (3.03 \pm 0.21)\times 10^{10}$ cm$^2/$s, where $D$ is the 
diffusion coefficient. The localization length for the samples with 
$d = 5$ cm is found below to be $\xi = 551 \pm 18$ cm.

In Fig. 1, we present the intensity distributions for two samples with 
$L/\xi \sim 0.1$ and 0.4. Calculations for diffusive waves in the absence 
of absorption have predicted that for $s_{ab} \gg g$, the intensity 
distribution falls as $exp(-2\sqrt{gs_{ab}})$. \cite{vR95} For the samples 
measured, we find $P(s_{ab}) \sim exp(-2\sqrt{\gamma s_{ab}})$ in the tail 
of the intensity distribution. The values of $\gamma$ obtained from a fit 
to the tail of the measured distributions is within $20 \%$ of the 
parameter $g^\prime = \Large \frac{2}{3 var(s_a)}\normalsize$,\cite{ms97} 
which equals $g$ in the absence of absorption . The fit of a stretched 
exponential to the distribution for $L/\xi = 0.4$ in the range of $s_{ab}$ 
from 10 to 18 is shown in Fig. 2 and gives $\gamma = 2.9$, which is close 
to value of $g^\prime$ of 3.06 for this sample.

The measured intensity distributions are compared to the transform of the 
measured transmission distributions \cite{ms97} for the corresponding 
samples using Eq.\ (\ref{psab}). The transforms shown as solid lines 
in Fig. 1, are in good agreement with the measured intensity distributions. 
A comparison of the moments of intensity and transmission is shown in Fig. 3. 
We find an increasing deviation of the ratios 
$<\!\!s_{ab}^n\!\!>/n!<\!\!s_a^n\!\!>$ from the value of unity expected 
from Eq.\ (\ref{moments}) as $n$ increases. We find, however, that agreement 
with Eq.\ (\ref{moments}) is dramatically improved when the moments are 
calculated using the asymptotic expressions in the diffusive limit for 
the intensity and transmission distributions beginning from the point at 
which the measured distribution have their first zero. The asymptotic 
expressions for the intensity distribution $exp(-2\sqrt{g^\prime s_{ab}})$ 
is substituted for the measured distribution for values of $s_{ab}$ between 
20 and 150, whereas the asymptotic exponential expression 
$exp(-g^\prime s_a)$ \cite{ms97,Kog95,vR95} is substituted for the measured 
transmission distribution for values of $s_{a}$ between 5 and 25. 
The improved agreement indicates that the extent to which the measured 
ratio of moments is in accord with Eq. (\ref{moments}) is largely limited 
by the range of intensity and transmission values measured, which depends 
on the number of configurations on which measurements were made.

Applying Eq.\ (\ref{moments}) to the second moments gives
\begin{equation}
\label{vars}
var(s_{ab}) = 2var(s_a) + 1.
\end{equation}
\noindent
The same expression can be obtained from a perturbation calculation up to 
order $1/g$ \cite{Kog95,Mello88}. Our measurements confirm the prediction 
of RMT that Eq.\ (\ref{vars}) is independent of the value of g and is 
correct up to order 1/N.

Intensity statistics at the localization threshold are studied in 
measurements at $L = 520$ cm and $d = 5$ cm and shown in Fig. 4a. 
Values  of $s_{ab}$ as large as 50 are observed. The distribution is seen 
in Fig. 4b to be nearly log-normal, in agreement with predictions for 
localized radiation \cite{Brw97,vLan96}. 

Using Eq.\ (\ref{psab}), we compare these measurements of the intensity 
distribution to random matrix calculations of the transmission 
distribution in the presence of absorption \cite{Brw97}. In order to 
compare the intensity distribution to theory, however, $\xi$ must be 
determined. Far from the localization threshold, in the absence of 
absorption and internal reflection, $var(s_a)$ and $\xi$ are related 
by \cite{Brw97,Kog95,vR95,Mello88}
\begin{equation}
\label{varsa}
var(s_a) = \frac{2L}{3\xi}.
\end{equation}
\noindent

To find $\xi$ in samples in which corrections due to absorption, 
localization and internal reflection cannot be ignored, we first obtain 
the intensity distribution for the equivalent samples without absorption 
using the measured spectra in our absorbing samples. The procedure used 
is based on the work by Yossefin \cite{Yos94} who proposed that for 
$\omega \tau_a \gg 1$ and when the dispersion is neglected the 
field in absorbing media differs from that in the absence of absorption 
only by a factor of $exp(t/2\tau_a$), where $\tau_a = {L_a}^{2}/D$ s is 
the exponential attenuation time due to absorption. We first obtain 
the time response $E(t)$ to a narrow gaussian pulse in time by Fourier 
transforming the product of the measured spectrum and a broad gaussian 
in the frequency domain. The time dependent field is then multiplied by 
$exp(t/2\tau_a$) and the so modified time spectra are transformed back 
into the frequency domain. The intensity distribution is then calculated 
using these field spectra. This procedure could be applied, however, 
only for $L \leq 150$. For longer samples the signal in time is buried 
in the noise level for large $t$ and thus a significant part of the 
field in the absence of absorption cannot be recovered. The intensity 
distributions corrected for absorption are in good agreement with 
transforms of the diffusive result for the distributions of total 
transmission calculated in Ref. \cite{Kog95,vR95} and give values for 
the parameter $g^\prime$ \cite{ms97} equal to the value of $g$, as 
expected in the absence of absorption. We also find that the average 
transmission obtained from the spectra corrected for absorption is 
consistent with the expected scaling as $(L + 2 z_b)^{-1}$, 
where $z_b$ is the diffusion extrapolation length due to internal 
reflection \cite{Lagd1}. These results confirm the ability of this 
approach to statistically eliminate absorption in the diffusive limit. 
The influence of internal reflection upon $var(s_a)$ in the absence of 
absorption can be accounted for by substituting $\tilde{L} = L + 2z_b$ 
for $L$ in Eq.\ (\ref{varsa}). We next account for the leading order 
correction to $var(s_a)$ due to nonlocal correlation. In the absence of 
absorption, the variance is increased by an additional factor of 
$(1 +\Large \frac{2\tilde{L}}{5\xi}\normalsize)$ \cite{Shap-pc,Brw-pc} 
to yield, 
\begin{equation}
\label{nabs}
var(s_a) = \frac{2\tilde{L}}{3\xi} + \frac{4\tilde{L}^2}{15\xi ^2}.
\end{equation}
\noindent
A fit of Eq.\ (\ref{nabs}) to the data corrected for absorption using 
$\xi$ and $z_b$ as fitting parameters gives $\xi = 551 \pm 18$ cm and 
$z_b = 5.25 \pm 0.31$ cm. The value of $z_b$ obtained is consistent 
with the values of this parameter for the same samples in the frequency 
range between 18 and 19 GHz. \cite{Gen93a} 

The dependence of $var(s_a)$ upon $L$ with and without absorption is 
shown in Fig. 5. The solid curve represents the result of the calculations 
in the diffusive regime ($L/\xi \ll 1$) by Brouwer \cite{Brw97} which 
account for absorption and the thin dashed curve shows the fit of 
Eq.\ (\ref{nabs}) to the data corrected for absorption. The values of 
$var(s_a)$ are calculated from $var(s_{ab})$ using Eq.\ (\ref{vars}). 
For lengths up to 200 cm, the result from the measurements, show sublinear 
behavior which is consistent with the results from total transmission 
measurements \cite{ms97} and the calculations in Ref. \cite{Brw97}. 
The deviation from the solid line increases for larger lengths and may 
reflect localization corrections that were not included in the theory. 
For strongly absorbing samples ($L \gg L_a$), Brouwer finds a log-normal 
distribution for the total transmission with 
$<lnT_a> = - L/L_a - 3L/4\xi - lnN$ and $var(lnT_a) = L/2\xi$. \cite{Brw97} 
Using thiese relations, we calculate the values of $var(s_a)$ in the regime 
of strong absorbtion. The result is shown by a thick dashed line. We note 
that this curve has a physical meaning for $L \gg L_a$ only. In this regime, 
it shows significant corrections due to localization in qualitative 
agreement with the results from the measurements. Thus, we can associate 
the increase of $\xi var(s_a)/\tilde{L}$ for these samples with the 
transition to localization.

We now compute $P(s_{ab})$ for the sample with $L = 520$ cm using the 
transmission distribution calculated in Ref. \cite{Brw97} and the values 
of $\xi$ and $L_a$ found here. The calculated intensity distribution is 
presented as the solid line in Fig. 4a and is in good agreement with the 
measurements. 

In conclusion, we find that the intensity distribution for $s_{ab} \gg g$ 
is described by a stretched exponential to power $1/2$ and that the 
distribution for $L \sim \xi$ is close to log-normal. We confirm 
experimentally the relationships obtained by Kogan and Kaveh between the 
moments and full distributions of intensity and total transmission. These 
relations unify the statistical description of local and spatially averaged 
transmittance quantities. Our measurements demonstrate that the statistics 
of wave transport is only marginally affected by absorption and that 
absorption does not substantially inhibit localization. The ability to 
reach the localization threshold using a quasi-one-dimensional sample is 
an extension to classical waves of the suggestion by Thouless that electrons 
will always be localized in sufficiently long wires at low temperatures. 
These results show that the variances of the intensity or transmission are 
reliable measures of the impact of localization upon transport in random 
media. 

We are pleased to acknowledge stimulating discussions with E. Kogan, P. W. 
Brouwer, B. Shapiro, R. Pnini, P. Sebbah, B. A. van Tiggelen, and A. Chabanov. 
We thank P. W. Brouwer for providing us with the results of his calculations. 
This work was supported by National Science Foundation Grant No. DMR9632789 
and by a PSC-CUNY award.

\pagebreak
\noindent
{\bf FIGURES}:\\

\noindent
Fig. 1. Probability distribution of intensity for samples with 
$L/\xi \sim 0.1$ and 0.4; the samples dimensions are (a) $d = 7.5$ cm, 
$L = 100$ cm, and (b) $d = 5.0$ cm, $L = 200$ cm, respectively. The solid 
lines represent distributions obtained from measured transmission distributions 
[9] using Eq. (2).\\

\noindent
Fig. 2. Fit of a stretched exponential of power 1/2 to the tail of the 
intensity distribution ($L/\xi \sim 0.4$).\\

\noindent
Fig. 3. Comparison between the moments of intensity and total transmission 
($L/\xi \sim 0.4$): \large $\bullet$ \normalsize moments obtained from the 
measurements, \large $\circ$ \normalsize moments calculated from the 
extended distributions.\\

\noindent
Fig. 4. Intensity distribution for the sample with $L = 520$ cm and $d = 5.0$ 
cm. The solid line in part (a) shows the distribution obtained from a 
transform of the total transmission distribution for this sample calculated 
using the expressions from Ref. [14]. The dashed line in part (b) represents 
a normal distribution.\\

\noindent
Fig. 5. Dependence of $var(s_a)$ upon $L$. The different symbols represent: 
\large $\bullet$ \normalsize results obtained from the measurements, 
\large $\circ$ \normalsize results from the data corrected for absorption. 
Note that the curves obtained using the theoretical expressions in Ref. [14] 
represent the ratio $\xi var(s_a)/L$, not $\xi var(s_a)/\tilde{L}$.\\


\begin{thebibliography}{99}
\bibitem{Shap86} B. Shapiro, Phys. Rev. B {\bf 34}, 4394 (1986)
\bibitem{Land70} R. Landauer, Philos. Mag. {\bf 21}, 863 (1970).
\bibitem{Goodman} J. W. Goodman, {\it Statistical Optics} (J. Wiley, 
New York, 1985).
\bibitem{Gar89} N. Garcia and A. Z. Genack, Phys. Rev. Lett. {\bf 63}, 
1678 (1989).
\bibitem{Dainty} J. C. Dainty, {\it Laser Speckle and Related Phenomena, 
Topics in Applied Physics v. 9}, ed. J. C. Dainty (Springer-Verlag, 
Berlin, 1984).
\bibitem{Gen93} A. Z. Genack and N. Garcia, Europhys. Lett. {\bf 21}, 753 
(1993). 
\bibitem{Shnerb} N. Shnerb and M. Kaveh, Phys. Rev. B {\bf 43}, 1279 (1991).
\bibitem{Thls77} D. J. Thouless, Phys. Rev. Lett. {\bf 39}, 1167 (1977).
\bibitem{ms97} M. Stoytchev and A. Z. Genack, Phys. Rev. Lett. {\bf 79}, 
309 (1997).
\bibitem{John84} S. John, Phys. Rev. Lett. {\bf 53}, 2169 (1984).
\bibitem{Gen91} A. Z. Genack and N. Garcia, Phys. Rev. Lett. {\bf 66}, 2064 
(1991).
\bibitem{Weav93} R. L. Weaver, Phys. Rev. B {\bf 47}, 1077 (1993).
\bibitem{Yos94} M. Yosefin, Europhys. Lett. {\bf 25}, 675 (1994).
\bibitem{Brw97} P. W. Brouwer, (Cond-matt/9711113) to be published.
\bibitem{Stone91} A. D. Stone, P. A. Mello, K. A. Muttalib, and J. L. Pichard, 
{\it Mesoscopic Phenomena in Solids}, eds. B. L. Altshuler, P. A. Lee, and R. 
A. Webb (Elsevier, 1991).
\bibitem{Kog95} E. Kogan and M. Kaveh, Phys. Rev. B {\bf 52}, R3813 (1995).
\bibitem{Kog93} E. Kogan, M. Kaveh, R. Baumgartner, and R. Berkovits, Phys. 
Rev. B {\bf 48}, 9404 (1993).
\bibitem{vR95} Th. M. Nieuwenhuizen and M. C. W. van Rossum, Phys. Rev. Lett. 
{\bf 74}, 2674 (1995).
\bibitem{vLan96} S. A. van Langen, P. W. Brouwer, and C. W. J. Beenakker, 
Phys. Rev. E {\bf 53}, 1344 (1996).
\bibitem{Gen93a} A. Z. Genack, N. Garcia, and A. A. Lisyansky, in {\it 
Photonic Band Gaps and Localization}, edited by C. M. Soukoulis (Plenum Press, 
New York, 1993).
\bibitem{Shap-pc} R. Pnini and B. Shapiro, private communication.
\bibitem{Mello88} P. A. Mello, E. Akkermans, and B. Shapiro, Phys. Rev. Lett. 
{\bf 61}, 495 (1988).
\bibitem{Lagd1} A. Lagendijk, R. Vreeker, and P. de Vries, Phys. Lett. A, 
{\bf 136}, 81 (1989). 
\bibitem{Brw-pc} P. W. Brouwer, private communication.
\end{thebibliography}
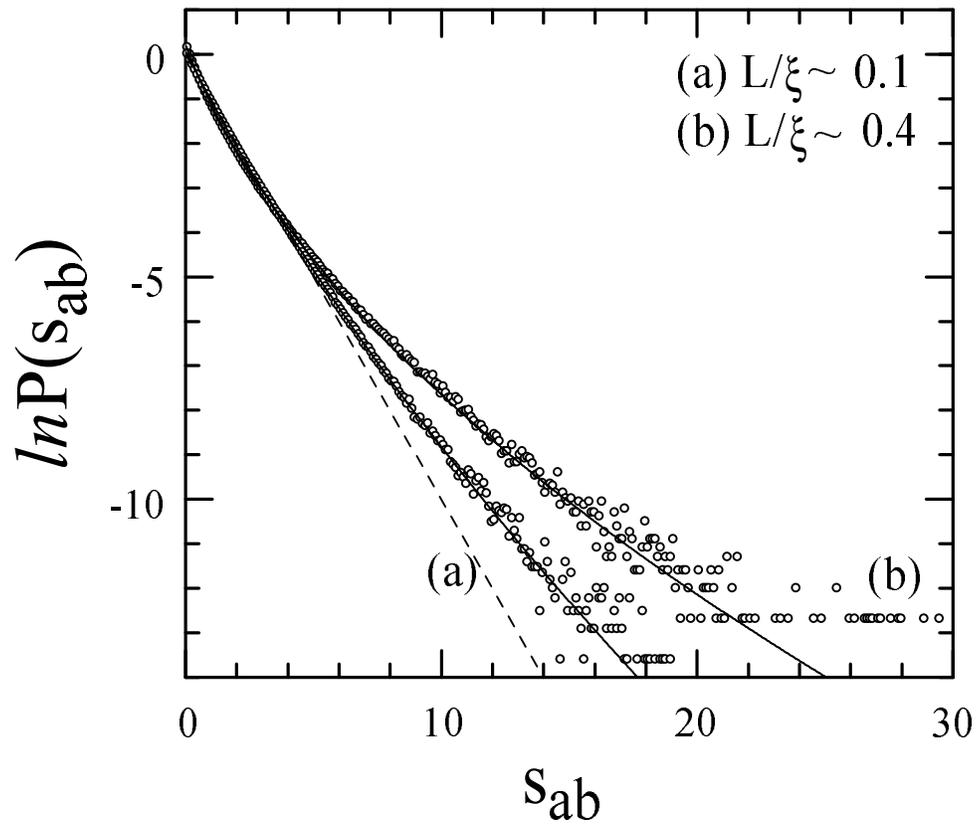
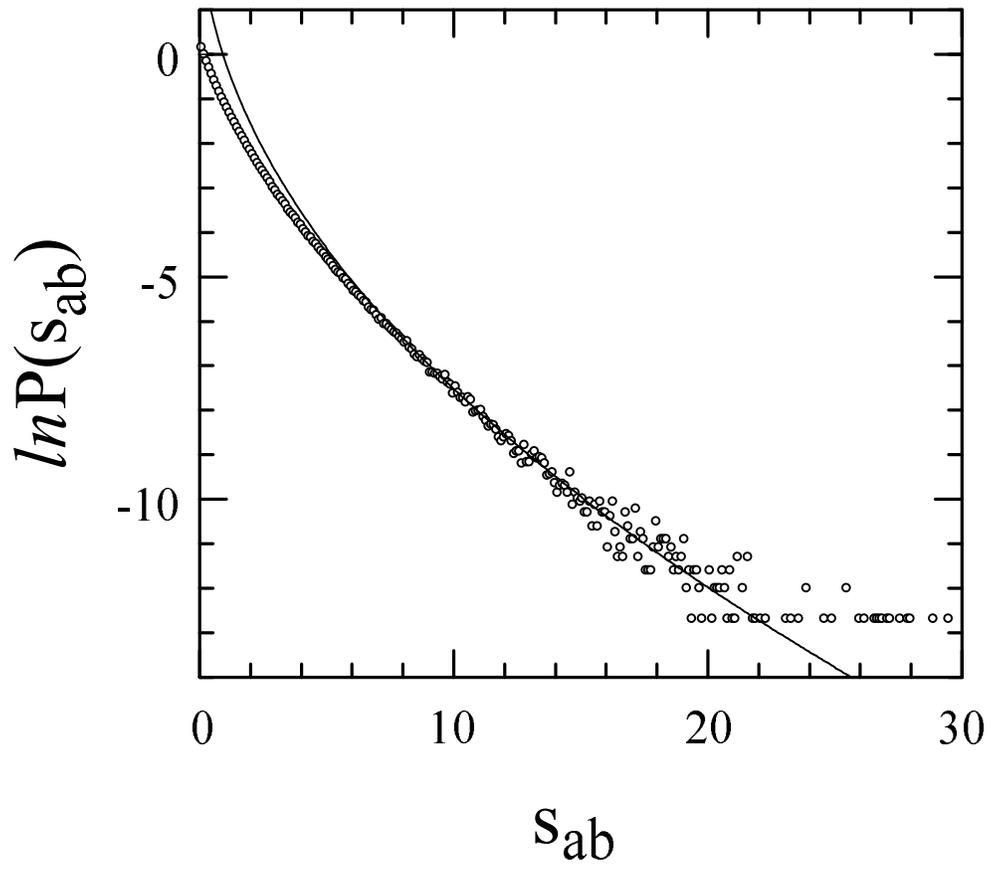
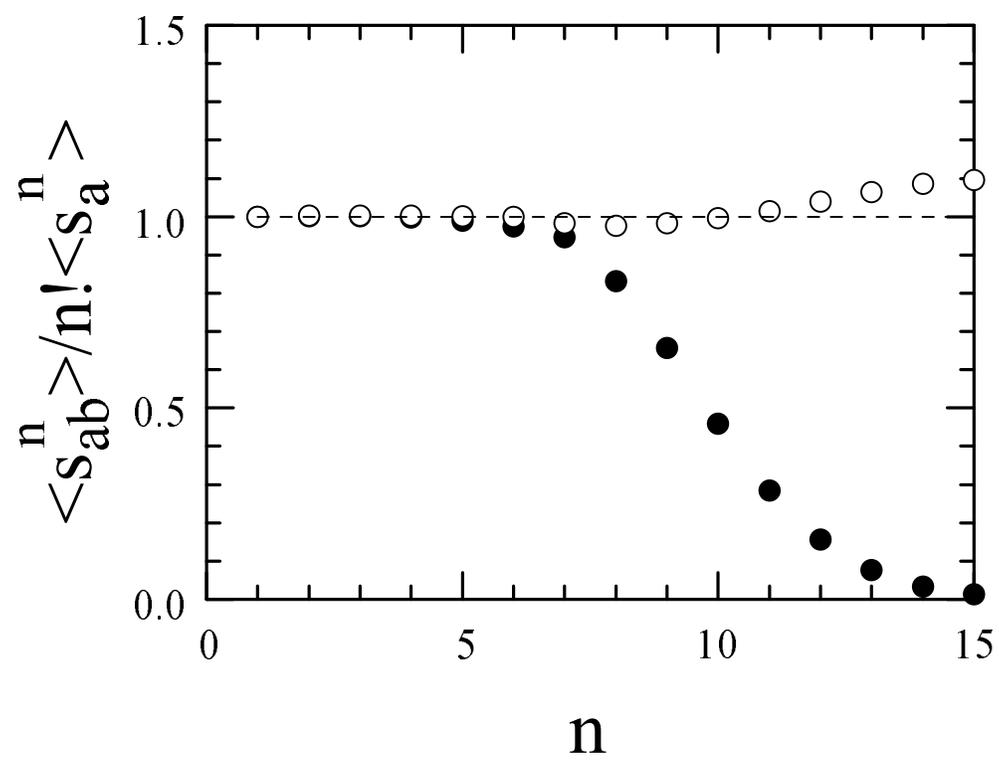
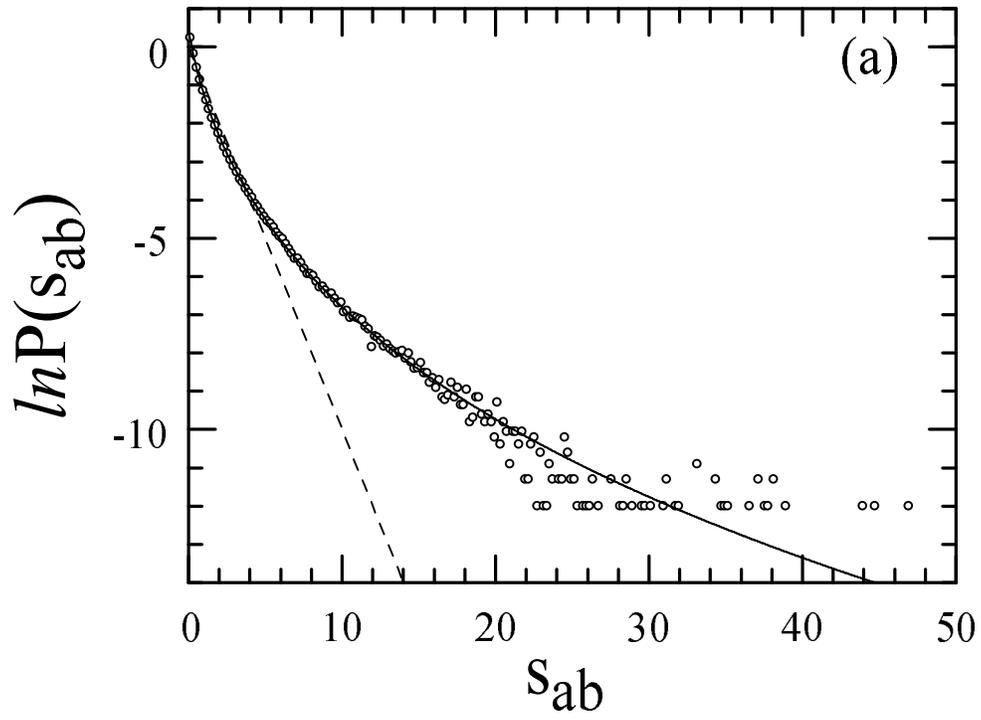
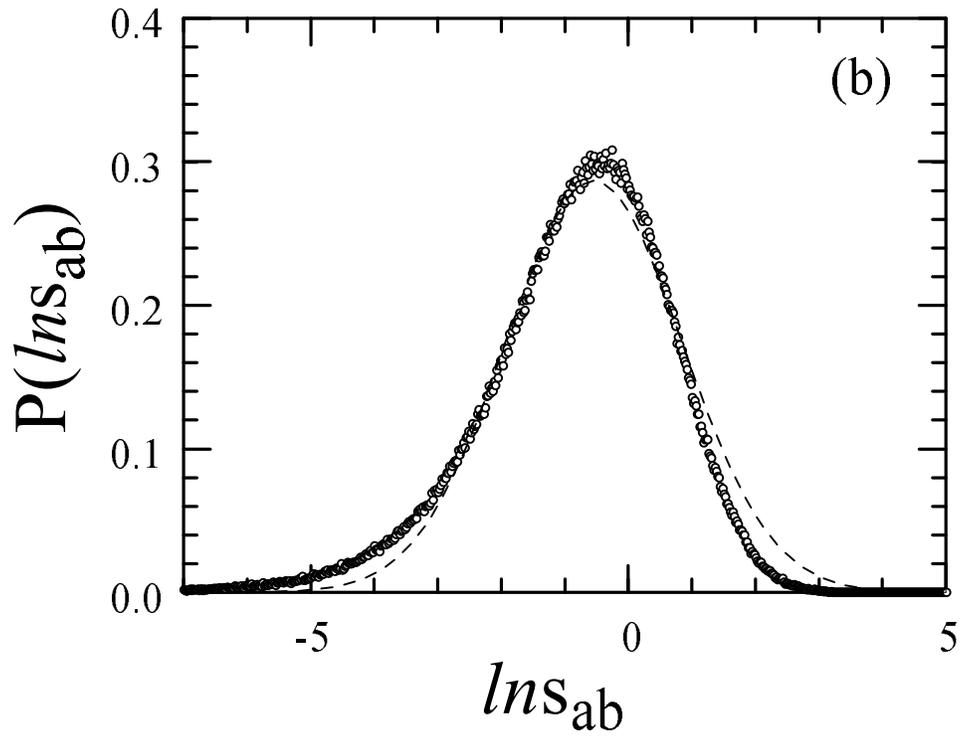
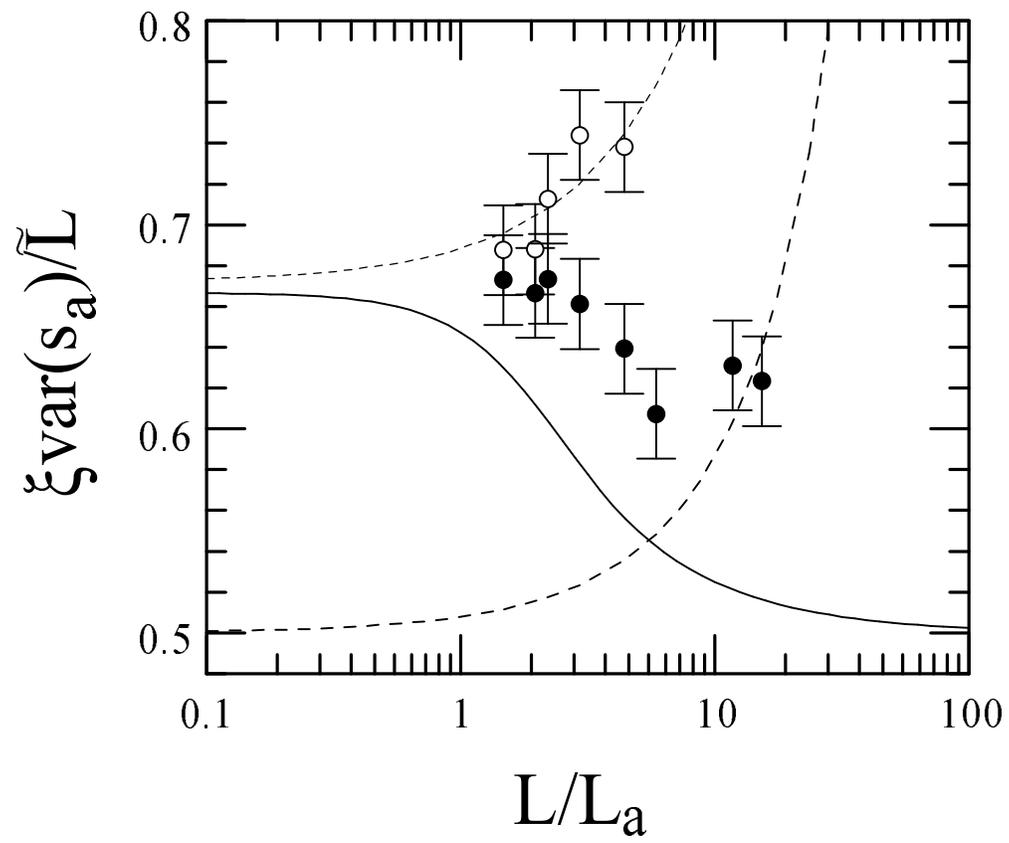
\end{document}